\documentclass[10pt,a4paper]{article}

\usepackage{graphicx,color}
\usepackage{amsmath,amsfonts}
\usepackage{tabularx}

\title{Detecting  dynamical changes in time series by using the Jensen Shannon Divergence}

\author{D. M. Mateos$^{1}$, L. Riveaud$^{2}$, P. W. Lamberti$^{2}$ \\
\footnotesize $^1$~{Neuroscience and Mental Health Programme, Division of Neurology, Hospital for Sick Children.}\\
\footnotesize{ Institute of Medical Science and Department of Paediatrics, University of Toronto, Toronto, Canada.} \\
\footnotesize $^2$~{Facultad de Matem\'atica, Astronom\'ia,  F\'isica y Computaci\'on (FaMAF), CONICET, and Universidad Nacional de C\'ordoba, C\'ordoba, Argentina} \\
\tt\footnotesize * mateosdiego@gmail.com
}

\begin{document}
\maketitle

\begin{abstract}

Most of the time series in nature are a mixture of signals with deterministic and random dynamics. Thus the distinction between these two characteristics becomes  important. Distinguishing between chaotic and aleatory signals is  difficult  because they have a common wide-band power spectrum, a delta-like autocorrelation function, and share other features as well. In general signals are presented as continuous records and require to be discretized for being analyzed. In this work we present different schemes for discretizing and for detection of dynamical changes in time series. One of the main motivations is to detect transition from chaotic regime to random regime. The tools used are originated in Information Theory. The schemes proposed are applied to simulated and real life signals, showing in all cases a high proficiency for detecting changes in the dynamics of the associated time series.

\end{abstract}


\section{Introduction}
\label{Introduction:sec}
In many areas of science  the dynamics of processes underlying the studied phenomena can be represented by means of finite time series, that can be interpreted as realizations of stochastic processes (SP) or as orbits of  deterministic processes. 
Although they are very different in nature, Wold demonstrated that under certain restrictions  there is a relationship between them \cite{Wold1938study}. His famous theorem  establishes  that a stationary stochastic process can be decomposed as a deterministic part (DP), which can be described accurately by a linear combination of its own past, and a second part represented  as moving average component of finite order. A different situation occurs when we have a non stationary  time series. In these situations it is not possible  to separate the series in a deterministic and in a stochastic part. Moreover, in cases in which the DP is chaotic, it is possible to find a DP which produces a time series that could be very difficult to distinguish of a  SP \cite{cencini2000chaos, sakai1980autocorrelations}. For these reasons, the issue of distinguishing between time series produced by deterministic chaos from noise has led to the development of different scheme of analysis.
This problem has already been treated with various techniques such as:   complexity-entropy plane \cite{Rosso2007, zunino2012distinguishing, olivares2012contrasting},   Lyapunov exponents \cite{brock1986distinguishing, cencini2000chaos, gao2006distinguishing} or applying neural networks \cite{elsner1992predicting}, among others.  Chaotic signals always produce time series with strong physical structure, unlike those originated in stochastic processes. These have a little structure depending on their correlations factors \cite{wu2004study}. The basic idea behind our schemes is to use quantifiers of how close or far are two time series. The quantifiers is the {\em Jensen--Shannon Divergence} (JSD), which was introduced as a measure of similarity between two probability distributions function (PDF) \cite{cover2012elements}.\\
Quantifiers based on information theory,  such as the JSD require to associate to a given time series a probability distribution (PD). Thus the determination of the most suitable PD is a fundamental problem, because of the PD  and the sample space  are inextricably linked. Many methods have been proposed to deal with this problem. Among others, we can mention binary symbolic dynamics \cite{mischaikow1999construction}, Fourier analysis \cite{powell1979spectral}, frequency counting \cite{rosso2009shakespeare}, wavelet transform \cite{rosso2001wavelet} or comparing the order of neighboring relative values \cite{Bandt2002}. In a similar approach, we  propose to explore a  way of mapping continuous-state sequences into  discrete-state sequences via the so-called {\em alphabetic  mapping}~\cite{yang2003linguistic}.  \\
Here we present two methods for analyzing time series. The first one  provides a relative distance between two different time series. It consists  in the  application of  the JSD  to a previously discretized  signal, through alphabetic  mapping. The quantity so evaluated will be called the \textit{alphabetic Jensen--Shannon divergence}  (aJSD). The second method detects the changes in the dynamics of a time series,  through an sliding window which run over the previously mapped signal. This window is divided in two parts. To each part we can assign the corresponding PD, which can be compared by using the JSD.
In both cases the change in the underlying dynamics of the time series can be detected by analyzing the changes in the value of the JSD at the neighborhood of the time when the change occurs.
Both methods were tested by using  well known chaotic and random  signals taken from the literature and  biophysical and mechanical real signals. \\
The structure of this work is the following: In  section \ref{JSD_Am_INTRO:sec} we give a brief introduction of the Jensen Shannon divergence, and we explained in detail the  alphabetic mapping. In section  \ref{aJSD_explanation} we introduce the aJSD. In section \ref{Characterization of Chaotic map and K-Noises} 
we give a brief characterization of the chaotic maps and colored noises. We  display the analysis of the chaotic and noisy series  in section \ref{Analysis of the aJSD between chaotic and random sequences}. In section \ref{aJSD_real_application}  we  illustrate  how the alphabetic Jensen Shannon divergence  can be used for real data extracted from biological and mechanical systems. We finish the paper by drawing up some conclusions.
%



\section{ Jensen-Shannon divergence for alphabetic mapping}
\label{JSD_Am_INTRO:sec}
\subsection{Jensen-Shannon Divergence}

The Jensen Shannon divergence (JSD) is a distance measure between probability distributions introduced  by C. Rao \cite{rao1987differential} and J. Lin \cite{lin1991divergence} as a symmetric version of the Kullback--Leibler divergence. I. Grosse and co-workers improved  a recursive segmentation algorithm for symbolic sequences based on the JSD  \cite{grosse2002analysis}. \\
Let $X$ be a discrete-states  variable 
$x_i$, $i=1,2,\ldots,N$ and let $P_1$ and $P_2$ be two probability
distributions for $X$, which we denote as $p_i^{(1)}=P_1(x_i)$ and
$p_i^{(2)}=P_2(x_i)$, with $0 \leq p_i^{(k)} \leq 1$ and
$\sum_{i=1}^{N} p_i^{(k)}=1 $  for all $i=1,2,\ldots,n$ and
$k=1,2$. If $\pi_1$ and $\pi_2$ denote the weights of $P_1$ and
$P_2$ respectively, with the restrictions $\pi_1 + \pi_2 =1$ and
$\pi_1,\pi_2 \geq 0$, the JSD is defined by
\begin{equation} 
D_{JS}(P_1,P_2) = H(\pi_1 P_1 +\pi_2 P_2) - (\pi_1 H(P_1) + \pi_2 H(P_2)) 
\label{DJS}
\end{equation} 
where
\begin{equation}
 H(P) = - C \sum_{i=1}^{N} p_i \; log (p_i)
\end{equation} 
is the Shannon entropy for  the probability distribution
$P$. We take $C=1/log(2)$ so that the entropy (and also
the JSD) is measured in bits.\\
It can be shown that the JSD is positive defined, symmetric and it is
zero if and only if $P_1=P_2$ \cite{cover2012elements}. Moreover JSD
is always well defined, it is the square of a metric \cite{endres2003new}  and it can be
generalized to compare an arbitrary number of distributions, in the following way: Let $P_1(x),...,P_M(x)$ be a set of probability distributions and let $\pi_1,...,\pi_M$ be a collection of non negative numbers  such that $\sum_{j=1}^{M} \pi_j=1 $. Then the JSD between  the probability distributions $P_j(x)$ with  $j=1,..,M$ is defined by 
\begin{equation}
D_{JS}\left( P_1,...,P_M\right)=H\left[\sum_{j=1}^{M} \pi_j P_j \right]- \sum_{j=1}^{M} \pi_j H \left[P_j \right].
\end{equation}


\subsection{Alphabetic Mapping }
\label{Alphabetic Mapping}
In  Nature it can be  found  discrete-state sequences, such as DNA sequences or those generated by logical circuits, but most of the `` the real life'' signals  are presented as continuous records. Thus to use the JSD first we have to discrete the data \footnote{  It is possible to calculate entropy for continuous states signals but the estimation of a differential entropy from the data is not an easy task \cite{beirlant1997nonparametric}.  }. There are many ways to overcome this requirement.  Here we use the \textit{alphabetic mapping} introduced by Yang  \cite{yang2003linguistic}.\\
For a given continuous series $ \{ x_t \} $, we can mapping this real-values series in a binary series easily, depending the relative values between two neighbor points $[x_{t};x_{t+1}]$ in the following way:
\[ s_{n} = \left\{ 
  \begin{array}{l l}
    0, & \quad \text{if $x_{t}\leq x_{t+1}$}\\
    1, & \quad \text{if $x_{t}>  x_{t+1}$}
  \end{array} \right.\]
$S$ is a binary sequence $S=\lbrace s_1,s_2,...,s_{N-1} \rbrace$ where $s_i=0$ or $1$. 
Consider two integers $d \geq 2$ and  $\tau \geq 1 $, let us then define a trajectory in the $d$-dimensional space as

\[ \mathbf{Y}_t \rightarrow [s_{t-(d-1)\tau}, ... , s_t ] \qquad t\geq (d-1)\tau \]

The vector $\mathbf{Y}^{(d,\tau)}_t$ is called \textit{``alphabetic word´´}, where $d$ is the \textit
{``embedding dimension´´} (the number of bit taken to create the word) and $\tau$ is called the \textit
{"delay"}. Taken's theorem gives conditions on $d$ and $\tau$, such that ${\mathbf{Y}^{d,{\tau}}_t}
$ preserves the dynamical properties of the full dynamic system (e.g, reconstruction of strange 
attractors) \cite{robinson2011dimensions, takens1981detecting}.

By shifting one data point at time, the algorithm produces a collection of bit-word $\lbrace \mathbf{Y}^{d,\tau} \rbrace$ over the whole series. Therefore, it is plausible that the occurrence of these bit-word reflects the underlying dynamics of the original time series. Different types of dynamics  produce different distributions on these $\mathbf{Y}^{d,\tau}$ series. We define $w^{d,\tau}_i$, as the symbol corresponding to the word $\mathbf{Y}^{d,\tau}_i$. From these  we construct a new series  $W^{d,\tau}=\{ w^{d,\tau} \}$ which quantify the  original series. The number of the different symbols (alphabet length) depends on the number of bit taken; in this case is $ 2^d $.

To give an example of the mapping we can consider the  series $ \{3.5, 4.8, 3.6, 2,1, 4.1, 3.7, 8.5, 10.4, 8.9 \}$, which has a corresponding the serie $S=\{1, 0, 0, 1, 0, 1, 1, 0  \}$. For the election of the parameter $d=4$ and $\tau=1$, the first word to appear is $\mathbf{Y}^{4,1}_1=(1,0,0,1)$, the second one is $\mathbf{Y}^{4,1}_2=(0,0,1,0)$, and the other three are  $\mathbf{Y}^{4,1}_3=(0,1,0,1)$, $\mathbf{Y}^{4,1}_4=(1,0,1,1)$ and $\mathbf{Y}^{4,1}_5=(0,1,1,0)$. 

Frequently  it is necessary to process  signals of two or more dimensions such as bi-dimensional chaotic maps, polysomnography, EEG, etc. The components of such signals are mostly coupled, given  signal values depending not only on the previous values but also on the values reached by the other signals. Therefore by making a one-dimensional analysis we can lost  some valuable information. Taking the idea discussed above for one-dimensional signals (1D), we have applied the same algorithm with a slight modification to analyze 2-dimensional (2D) signals without loosing information.\\
For a given continuous 2D series $ \mathbf{X}_{t} = (x (t), y (t))$, we assign a 1D string, by the relative values between the two components vector at each time $t$,  $[x_t; y_t]$ in the following way:
\[S_t = \left\{\begin{array}{lll}
0  & \mbox{si} & x_t \geq y_{t}\\[2mm]
1 &  \mbox{si} & x_t <  y_{t}
\end{array} \right.\]


\section{JSD combined with the alphabetic mapping }
\label{aJSD_explanation}

Once discretized the signals  we can approximate  the PD  by  the frequency of occurrence of the symbols $W^{d,\tau}$. From this PD, we can develop the following analysis schemes:

\begin{itemize}
\item \textbf{Distance between two signals:}
Given two different sequences, for example, a chaotic and a random sequences, we  map each one  
by using the methodos explained in section \ref{Alphabetic Mapping}. This gives us two sets of symbols $
W^{(d,\tau)}=\{ w^{(d,\tau)} \}$ and $\widetilde{W}^{(d,\tau)}=\{ \widetilde{w}^{(d,\tau)} \}$. We calculate the frequency appearance of 
the symbols for both sequences  $P^W = P(W^{d,\tau})$ and  $P^{\widetilde{W}}=P(\widetilde{W}^{d,\tau})$.  Finally, we calculate the Jensen Shannon divergence  between these two distributions $D_{JS}^{W,\widetilde{W}}=D_{JS}(P^{W} \lvert P^{\widetilde{W}})$. \\

\item \textbf{Sliding Window:}
We introduce an sliding window that moves over the symbolic sequence corresponding to the original signal. The window has a width $ \Delta> 0 $ and the position $ k $ (referring the position of the center of the window over the sequence). For each position $ k $, we can divide the window in two sub windows, one to the left and the other to the right of the position $k$. For both windows, we evaluate the frequency of occurrence of symbolic patterns: to the right $P(W^{d,\tau}_{r})=P^W_{r}$ and to the left $P(W^{d,\tau}_{l})=P^w_{l}$. Finally we evaluate the associate JSD,  $D_{JS}(k)=D_{JS} (P^{W}_{r} \lvert P^W_{l})$ as a function of the pointer position $k$. The position where the maximum value of the JSD occurs, $D_{JS_{max}}=max[D_{JS}(k)]$,  it is interpreted as the place where a significant change in the probability distribution patterns $W^{(d,\tau)}$, has occurred.  This change can be associated to a variation in the statistical properties of the original signal.
The only restriction in the election of window's width is that $ \Delta $ must be greater than the number of possible patterns generated by the alphabetic mapping $ ( \Delta >> 2 ^ d) $ . 

\end{itemize}
Two sequences with the same statistical properties, should lead to identical probability distributions and therefore, the divergence between them should take a very small value, close to zero but non zero. This fact is due to the statistical fluctuations. The estimators for probability distributions corresponding to sequences must  be constructed and then the fluctuations of this construction will yield to JSD values greater than zero. To address this problem Grosse et al. \cite{grosse2002analysis}  introduced a quantity called "significance" which allows  to see if the values reached by the JSD are greater than the statistical fluctuations. This amount depends on the sequence length and the size of the alphabet of symbols used in the representation of the sequence. An expression for the significance value has presented in the reference  \cite{grosse2002analysis}. A limitation of that expression is that it is valid only for an alphabet with no more than five symbols.  Therefore we must  modify the criteria introduced by Grosse et al, to identify values of the JSD that are genuinely  above statistical fluctuations. To do that, we proposed to calculate the aJSD on a set of $Ns=10^5$ ensembles generated for each signal using the same parameters but with different initial conditions. After quantifying the signals we have $Ns$ differents sequences $(W^{d,\tau}_1,....,W^{d,\tau}_{Ns})$.

The first step is to calculate the aJSD between all sequences belonging to the same group of segments, that we call ``auto-aJSD´´,

 \[ D_{JS}^{W}=D_{JS}(P^{W_i}  ~ \lvert ~P^{W_j}),~ i,j=1,...,Ns ,~ for ~ i\neq  j \]

Then we evaluate the  average value $ \mu^W= \langle D_{JS}^{W} \rangle$ over all the sets of sequences with its respective standard deviation ($ \sigma^{W}= \langle (D_{JS}(P^{W_i} \lvert P^{W_j})-\mu^{W} )^2\rangle^{1/2} $).
The next step is the same, but using the two group of signals  to be compared. For all the aJSD values resulting  from all the signal we take the average   $\mu^{W,\widetilde{W}}= \langle D_{JS}(P^{W_i} \lvert P^{\widetilde{W_i})} \rangle $, with its respective standard deviation ($ \sigma^{W,\widetilde{W}}= \langle (D_{JS}(P^{W_i} \lvert P^{\widetilde{W_i}})-\mu^{W,\widetilde{W}} )^2\rangle^{1/2} $).

 Finally two sequences are different (in the statistical sense) if the inequality:

\[ \mu^{W,\widetilde{W}} - \sigma^{W,\widetilde{W}} \geq  \text{max} [ \mu^{W} + \sigma^{W}, \mu^{\widetilde{W}} + \sigma^{\widetilde{W}}] \] 

is satisfied. If the values of the aJSD do not pass this criteria we say that the two signals are not statistically distinguishable one from each other.


\section{Characterization of chaotic maps and colored noises}
\label{Characterization of Chaotic map and K-Noises}

To test our scheme of analysis we used  sequences extracted from the bibliography. We use  18 chaotic maps and 5 colored noises, that we describe briefly. 
\subsection {Chaotic maps}

We consider 18 chaotic maps which were taken from the reference \cite{sprott2003chaos}. They can be grouped as follows:

\begin{enumerate}

\item \textbf{1D chaotic maps:} also call non--inverted maps. They are dynamical sequences for wich the image has more than one pre-image and in each interaction  a loss of information occurs,  generating in this way a chaotic system  \cite{sprott2003chaos}. \\
\begin{itemize}
\item  the lineal congruential generator \cite{knuth1973sorting}, 
\item the gaussian map \cite{steeb1992chaos}, 
\item the logistic map \cite{may1976simple}, 
\item the Pinchers map \cite{potapov2000robust}, 
\item the Ricker's population model \cite{ricker1954stock}, 
\item the sine circle \cite{arnol1961small}, 
\item the sine map \cite{strogatz2001nonlinear}, 
\item the Spencer map \cite{shaw2000strange} and 
\item the tent map \cite{hirsch2004differential}.
\end{itemize}

The maps that have been used are :
\begin{itemize}
\item  the H\'enon map  \cite{henon1976two},
\item the Lonzi map,  
\item the delayed logistic map  \cite{aronson1982bifurcations},
\item the tinkerbell map  \cite{nusse1998dynamics},
\item the dissipative standard map \cite{schmidt1985dissipative},
\item the Arnold's cat map \cite{arnold1968ergodic},
\item the chaotic web map \cite{chernikov1988chaos},
\item the Chirikov standard map \cite{chirikov1979universal},
\item the gingerbreadman map \cite{devaney1984piecewise} and 
\item the H\'enon area-preserving quadratic map \cite{henon1976two}.
\end{itemize}

\end{enumerate}

\subsection {Colored noises}
Noises with a power spectrum that varies with frequency are called \textit{colored noises}. 
There are many types of noises, depending on the shape of the power spectrum and the distribution of values. The noise power spectrum often varies with frequency as $1/f^\alpha$ (some time called \textit{Hurst noise}).  \textit{White noise} correspond to  $\alpha=0$.
To generate all noises  we used the algorithm described in \cite{Rosso2007,LarrondoProg}


\section{ Distinguishing between chaotic and random sequences}
\label{Analysis of the aJSD between chaotic and random sequences}

In this section we present the results obtained from evaluating the aJSD between chaotic maps and colored noises. We apply the two methods described in section \ref{JSD_Am_INTRO:sec}. First we create an aJSD distance matrix between chaotic and noise signals and between chaotic and chaotic maps. In the second place we merge two different signals (i.e. one chaotic and one noisy)  and through the aJSD sliding window method  it is possible to detect  the signal changes from one regime to another.

\subsection{ Distance matrix between sequences}
\label{Distance matrix between sequences}
\begin{figure}
  \begin{center}
    \includegraphics[scale=0.7]{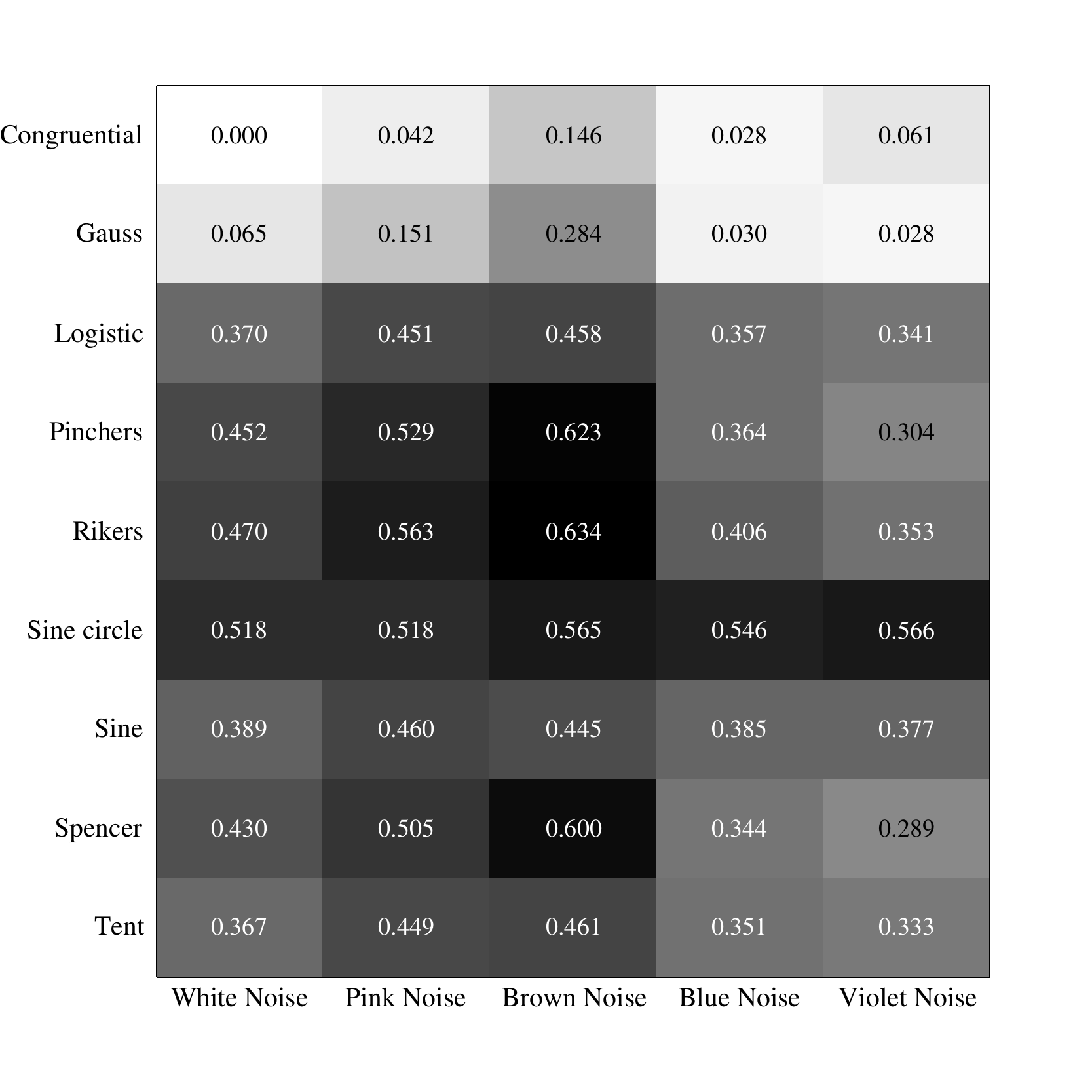} 
    \caption{aJSD distance matrix between 1D chaotic sequences and  colored signals. The values chosen of the parameters $d$ and tau are:  $d = 8$ and $\tau = 1$. It can be observed that there is a good discrimination between chaotic sequences and noise. aJSD values vary according to the correlation noise and the dynamics of chaotic series. The value 0.00 means that the aJSD has not exceeded the significance criterion set out in section \ref{aJSD_explanation} so that the two sequences are indistinguishable from each other. Similar results were found for parameters $8<d<12$ and $\tau=1$, showing that the method is robust respect the election parameter $d$.}
    \label{fig:CN_matrix_1D}
  \end{center}
\end{figure}

For each type of process  explained in the previous section we generate $N_s=10^ 6$ time series of $L _s = 10^6$ data points with identical parameters  \cite{sprott2003chaos} and a random initialation. We compute a distances matrix between  chaotic and colored noises, using the   significance criterion for the  aJSD values as explained in section \ref{aJSD_explanation}. For the discretization of the signals we used the parameters $\tau = 1$ and $8<d<12$.  Fig. \ref{fig:CN_matrix_1D} displays the matrix for the corresponding parameters  $d = 8$ and $\tau = 1$. For different embedding dimension, we obtained similar results. 

In the case of the aJSD distances matrix  corresponding to chaos-noise (Fig: \ref{fig:CN_matrix_1D}) we can observe that most of chaotic maps are distinguishable from the different types of colorated noises. The number in the boxes represent the values of the aJSD. Lower is this value, more similar are the sequences. 

Only  for the particular case of the lineal congruential generator map (LCG) and white noise (WN), the aJSD value does not pass the significance criterion. An interpretation of this is that the LCG map is an example of a random number generator passing the Miller--Rabin test \cite{rabin1980probabilistic}. Therefore the distribution of words $\{W^{(8,1)}\}$ corresponding to LCG and WN are similar.\\

\begin{figure}
  \begin{center}
    \includegraphics[scale=0.7]{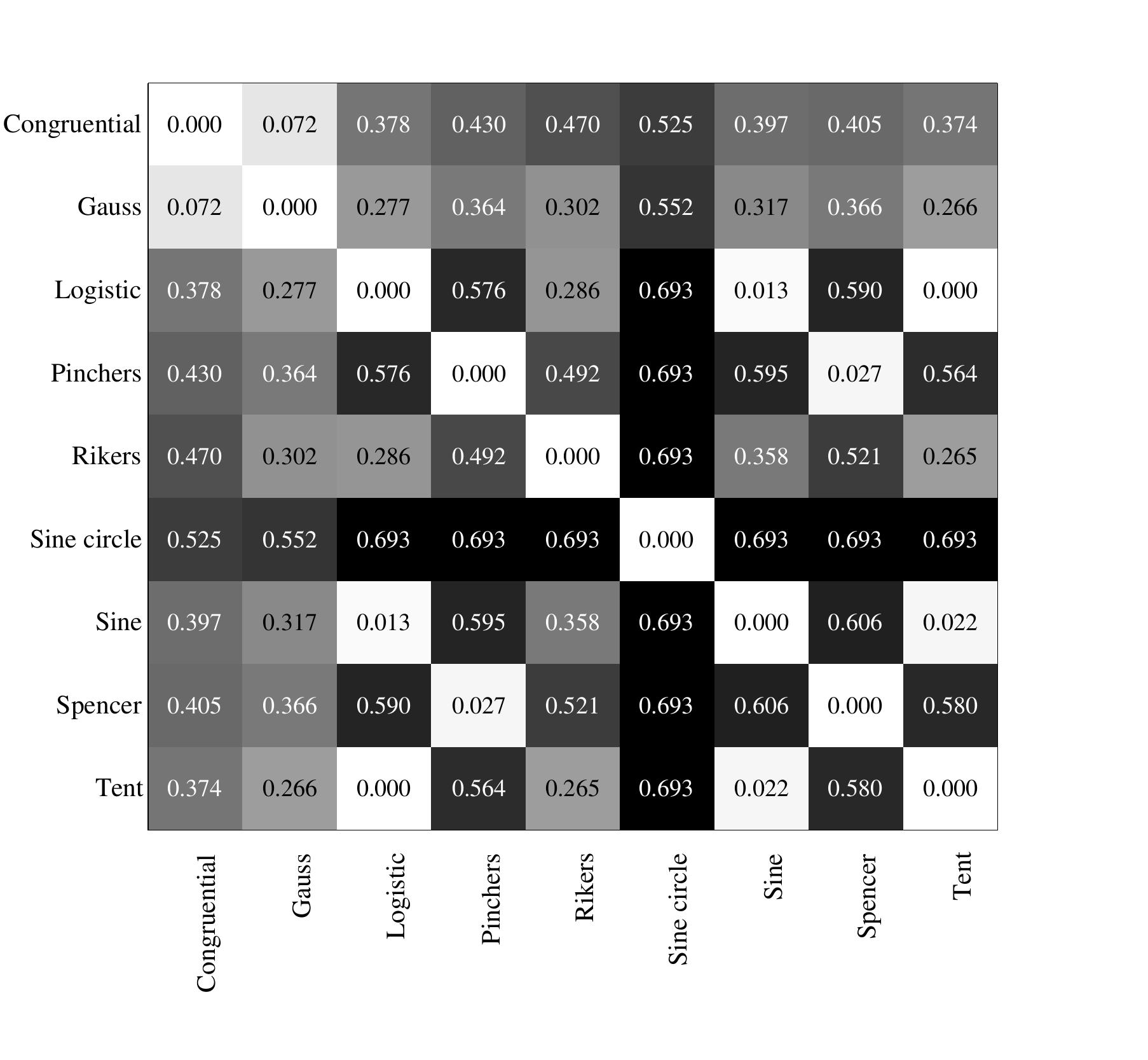}
    \caption{aJSD distance matrix between 1D chaotic sequences. The values chosen for the parameters $d$ and $\tau$ are: $d = 8$ and $\tau = 1$. It can be observed  a good discrimination between the different chaotic sequences. The absolute values of aJSD vary according to the similarity between the phase diagrams of the chaotic sequences, giving information of the similarity about the dynamics of such sequence. Similar dynamics give similar sequences. Similar results were found for parameters $8<d<12$ and $\tau=1$}
        \label{fig:CC_matrix_1D}
  \end{center}
\end{figure}

\begin{figure}
  \begin{center}
    \includegraphics[scale=0.7]{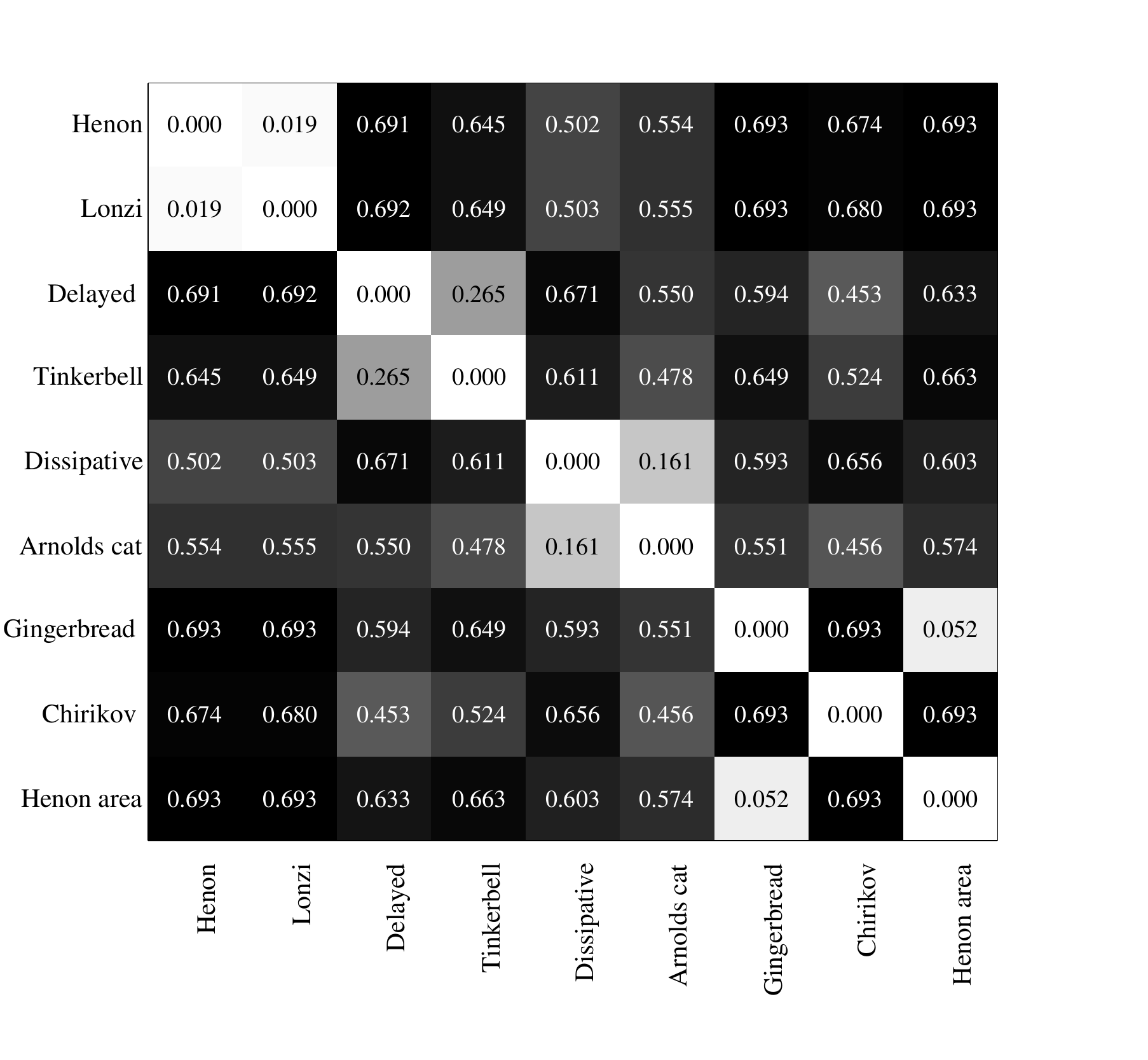}
    \caption{aJSD distance matrix between 2D chaotic sequences. The values chosen for the parameters $d$ and $\tau$ are: $d = 8$ and $\tau = 1$. The matrix shows  aJSD values is a good discrimination between the different chaotic sequences. As happen in the 1D chaotic maps, when more similar are the phase space , lower is the value of aJSD. Similar results were found for parameters $8<d<12$ and $\tau=1$}
    \label{fig:CC_matrix_2D}
  \end{center}
  
\end{figure}

The same evaluation have been done between chaotic maps.  Figure \ref{fig:CC_matrix_1D} displays the corresponding distance matrix. As it can be seen all the aJSD values are above significance criterion, showing our method to be adequated to distinguish between different types of chaos. A more detailed analysis shows a strong relationship between the values of aJSD and  the phase diagram of the chaotic maps. For example, the logistic map and the  tent map  have similar phase diagrams \footnote{ All phase diagrams belong to chaotic time series used in this work are shown in the Sprott book  (Apendix~\cite{sprott2003chaos})}, and the value of the distance between both sequences is very low. The same behaviour can be found between the  Pincher's map and the Spencer map. \\
The same analysis was performed for 2D chaotic  series. In this case we use the two-dimensional assignment  described in section  \ref{Alphabetic Mapping} and where the parameters $8 <d< 12$ and $\tau = 1$ were chosen .
Fig \ref{fig:CC_matrix_2D} shows the corresponding  aJSD distance matrix for the  parameters $d = 8$ and $\tau = 1$. As observed for one-dimensional chaotic maps, all distances over passed the significance criteria, being all maps distinguishable one from each other. 
There exists an strong correspondence between the values of the aJSD and the similarity (or dissimilarity) of the topology of the phase diagrams of the maps.  
The aJSD values ​​decrease as  the topology of the phase diagram tend to be similar,  can see it in the case of H\'enon map and the Lonzi map. Conversely when two maps are topographical different, such H\'enon map preserver area and Chirikov map, the aJSD increas. 
Different embedding dimensions  changes the absolute value of the aJSD but  the relative value between the elements of the distance matrix remain unaltered.\\

Let us recall that in our scheme the aJSD measure the distance between  the PDF associated with the set of words $ \{W^{(d, \tau)}\}$ and $\{ \widetilde{W}^{(d, \tau)}\}$.  As a consequence of the Taken's theorem \cite{takens1981detecting}, these sets of words are in correspondence with certain aspects of the phase space of each original signal. Series which have a similar dynamics, have similar phase space and PDF comparable giving  low values of the aJSD.\\

\subsection{Detection changes in time series}
\label{The aJSD as   changes signal detector }

\begin{figure}
  \begin{center}
    \includegraphics[scale=0.6]{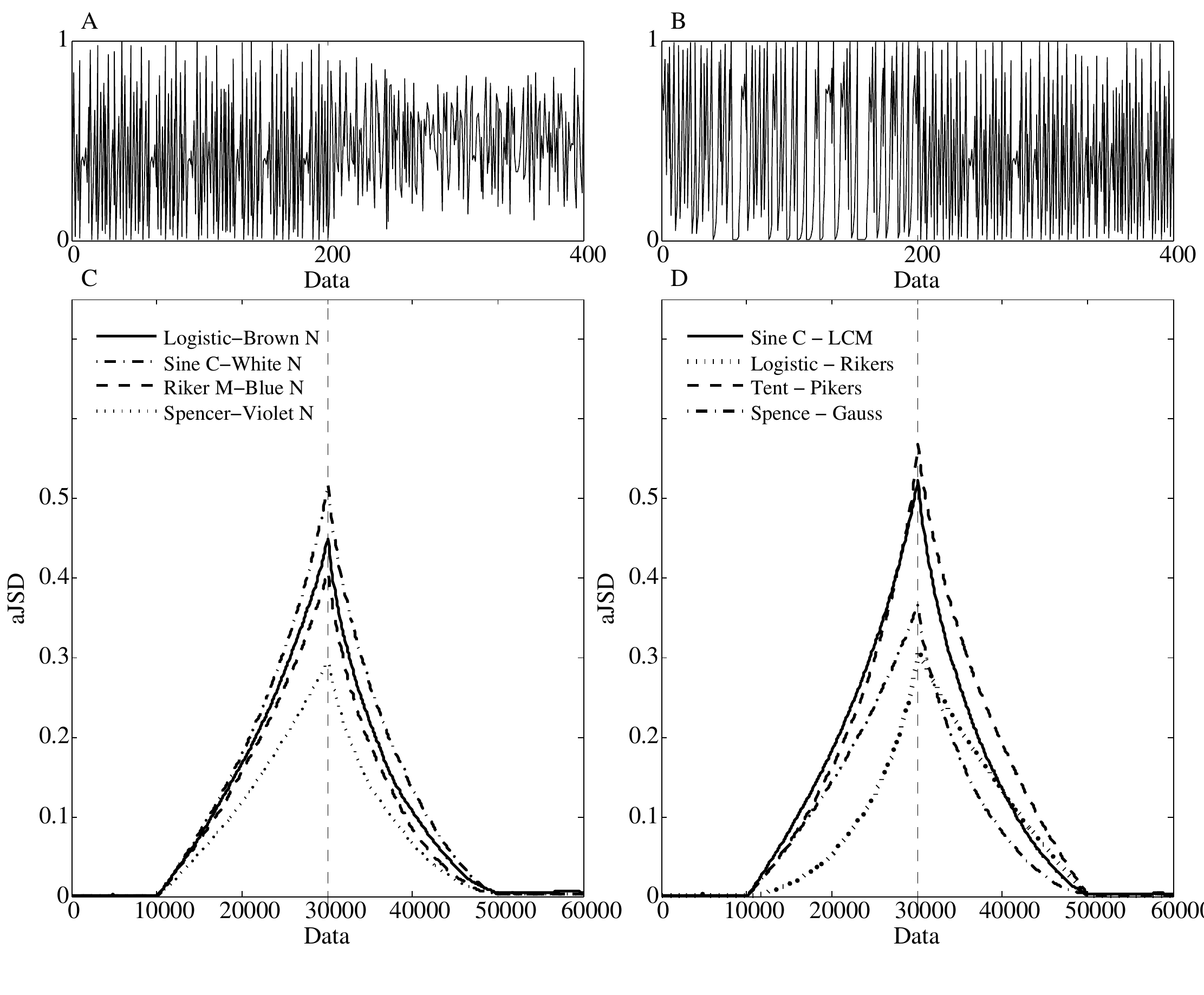} 
    \caption{(A) Example of a combined chaos-noise  signal (Rikers maps - Blue noise). The signals were  merged at the meddle position (point=200). (B) idem A but using a chaos-chaos combined signals (Sine Circle - Linear congruential Map). (C) aJSD calculation using the running window method, applied to four composed chaotic and random sequences. The values chosen for the parameters $d$ , $\tau$ and $\Delta$ are: $d = 8$, $\tau = 1$ and $\Delta = 40000$, . (D) Idem C applied  to  four  composed  different chaotic signals. In both plots,  the maximum aJSD value is reached  at the point where the two different sequences were merged (dotted vertical line)}
\label{fig:aJSD_wind_1D}
  \end{center}
\end{figure}

Here we use the proposed sliding window scheme for detecting changes in a signal. For this purpose we used  two different signals of equal length $L_{x_1}=L_{x_2}=5.10^4$ symbols each one, which are merged  in a single  sequence, where the signal $x_1$ is a chaotic one and $x_2$ is a random one, or two different chaotic sequences. Examples of two combined normalized sequences are plotted in Figures. \ref{fig:aJSD_wind_1D} A and B. 
Figure \ref{fig:aJSD_wind_1D} shows the results obtained by applying the aJSD to four combinations of chaos-noise sequences. The aJSD achieves its maximum value exactly at the merging point of the two  sequences, which is marked with a dotted vertical line. Similar results   were observed in the case of sequences generated by chaotic processes. In these cases the aJSD value reaches  several orders of magnitude higher than those corresponding to a single stationary sequence.

%
%


\section{Application of the aJSD to real data}
\label{aJSD_real_application}

Now we test our proposed schemes for a group of real world signals. The first one is a set of ECG signals   on which we calculated the distance matrix of aJSD between groups of patients with different cardiac pathologies. In the second application we use our methods to detected the alignment of the axis of an electric motor.

\subsection{Distinction between groups of patients with heart diseases}
\label{Distinction between groups of patients with heart problems}

\begin{figure}
  \begin{center}
    \includegraphics[scale=0.5]{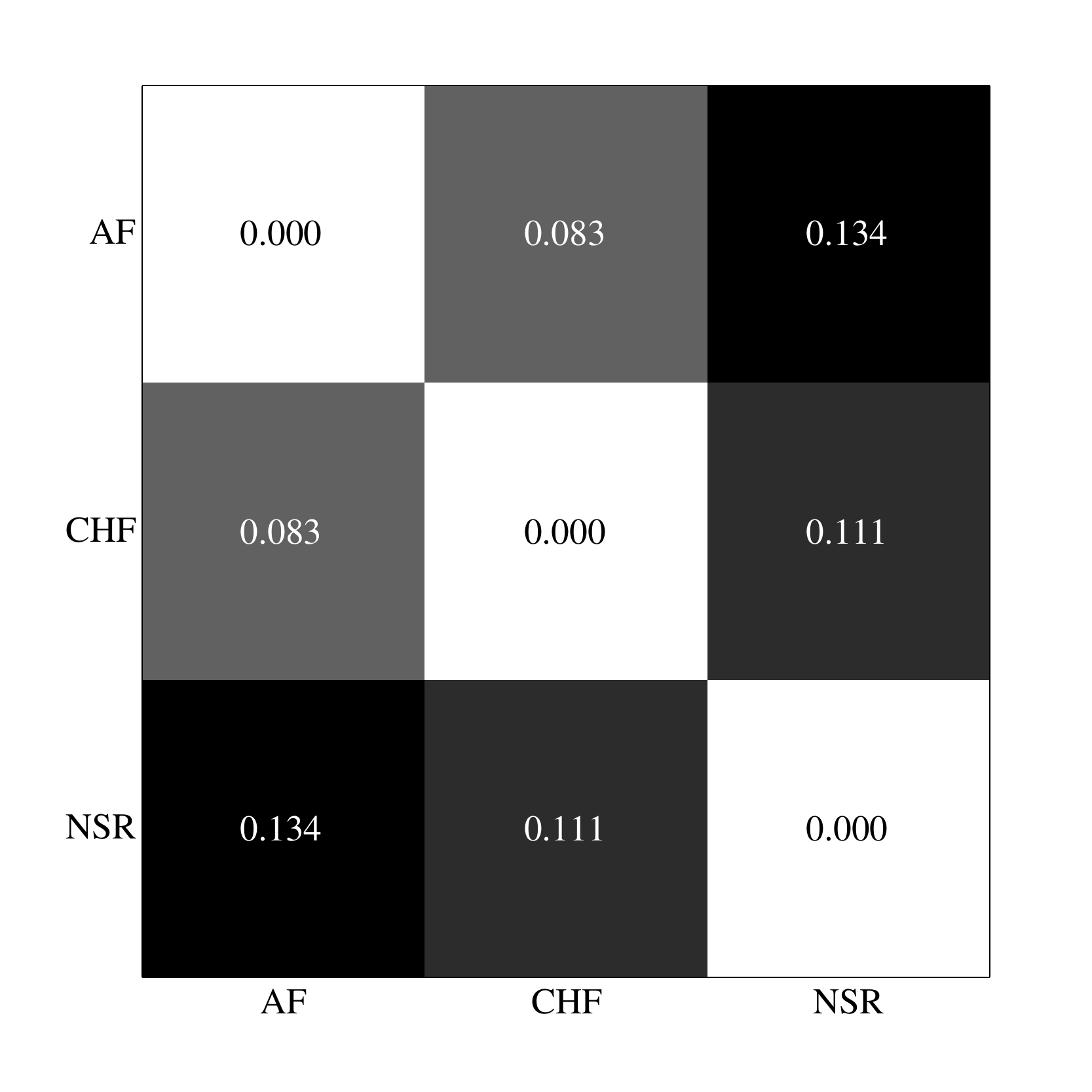} 
    \caption{  aJSD distance matrix for three groups of 5 patients each one: the normal sinus rhythm (NSR) group, the congestive heart failure (CHF) group and the atrial fibrillation (AF) group. The aJSD can discriminate between this tree groups.The parameter used in this matrix was $d=8$ and $\tau=1$. The same results were obtained for higher embedding dimensions ($8 \leq d \leq 12$) .}
    \label{fig:aJSD_ECG}
  \end{center}
\end{figure}

In this study case the signals correspond to the time intervals between heartbeats (BBI) on 15 patients. The patients were grouped into 3 set of 5 patients each one. The first group consists of healthy persons  with normal sinus rhythm (NSR); the second set contains patients suffering from congestive heart failure (CHF), and the third one is composed  of patients suffering  atrial fibrillation (AF). These set of data are freely available at: www.physionet.org/challenge/chaos/.\\
 Each series represents a record of  24 hours (approximately 100,000 intervals). The analysed records do not have any previous filter. Figure \ref{fig:aJSD_ECG} plots the  aJSD distance matrix  among the three sets, for the parameters ( $d = 8$ and $\tau = 1$). The aJSD can discriminate between the members of the control group  and the pathological groups, being greater the distance between the members of the pathological groups and the members of the control group. The frequency  time between heartbeats for the patients belonging to the control group,  remains virtually unchanged, while for patients with pathologies the interbeats time is altered. The difference between heartbeats is captured through the alphabetical mapping, giving different patterns distributions of $W^{d,\tau}$ for each group. \\

\subsection{Misalignment detection of an electric motor}
\label{Misalignment detection of an electric motor}

\begin{figure}[htbp]  
  \begin{center}
    \includegraphics[scale=0.4]{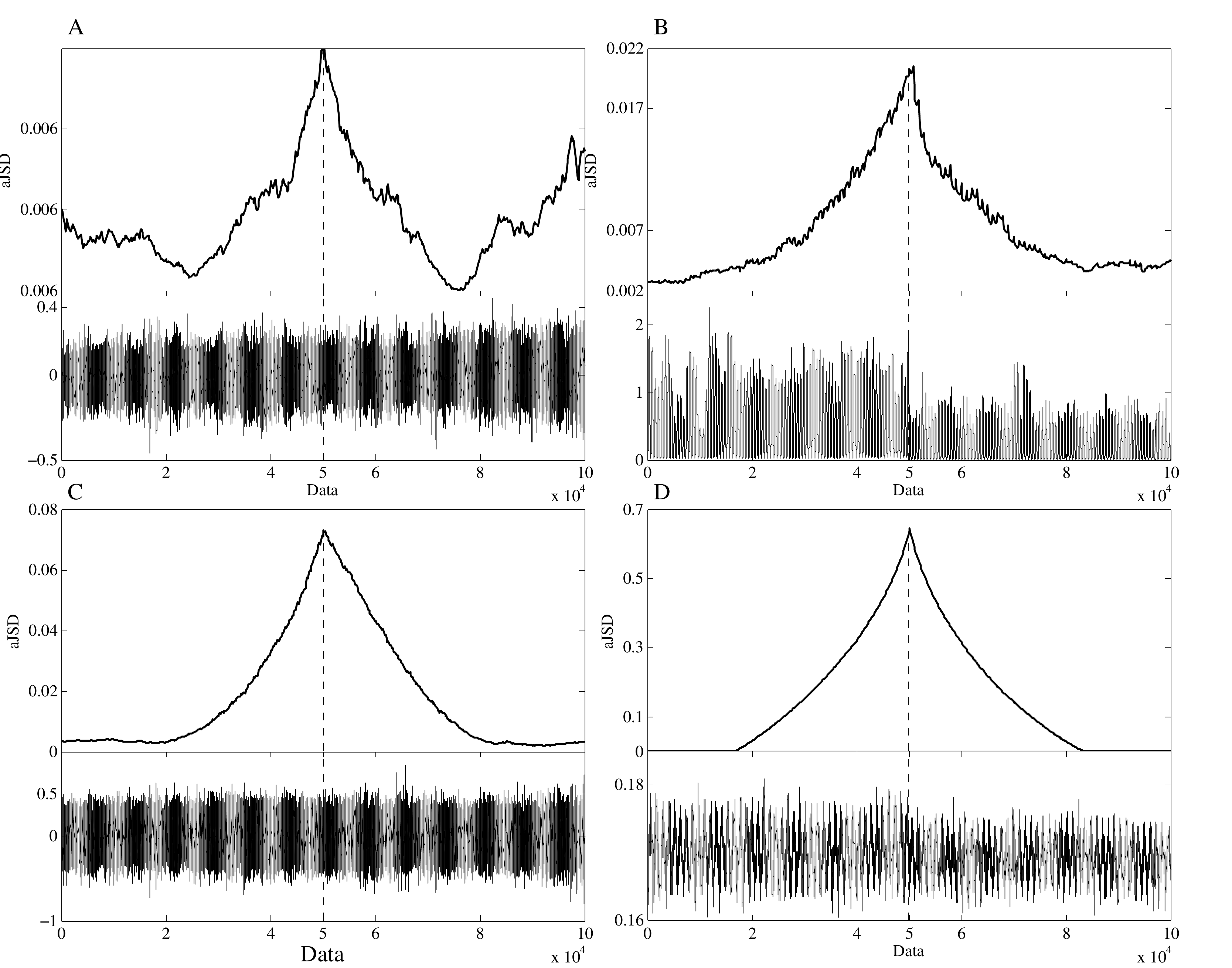}
    
    \caption{aJSD analysis using a sliding windows, corresponding to the four different signals taken on the axis of an electric motor. (A) Signal of a capacitive accelerometer mounted on the $z$ axis of the motor bearing. (B) An optical signal incremental rotary encoder  drum with 145 pulses per revolution. (C) Signal generated by a piezoelectric accelerometer mounted on the same place as A. (D) Signal engine load rate. All signals were analysed choosing the parameters $d = 8$, $\tau = 1$, $\Delta = 40000$.}
  \label{fig:aJSD_wind_GIPSA}
  \end{center}
\end{figure}

The second example corresponds to a record taken from the axis movement of an electric motor. The first signal corresponds  to the vibration measure using a capacitive accelerometer, mounted along the $z$ axis on the  motor bearing (Figure~\ref{fig:aJSD_wind_GIPSA}A). The second one, is a signal taken from an optical incremental  rotary encoder drum, with about 145 pulses per revolution as shown in Figure (~\ref{fig:aJSD_wind_GIPSA}B). The third is a signal generated by a piezoelectric accelerometer mounted in the same place that the capacitive signal, which is plotted in Figure (~\ref{fig:aJSD_wind_GIPSA}C). The last signal is the  rate of the engine load and depicted in  Figure ~ (\ref{fig:aJSD_wind_GIPSA}D) \footnote{It is related to the frequency of rotation of the shaft by means of a "Keyphaser"} . The data were obtained with a sampling frequency of $25.10^3$ Hz without any preprocessing. For more technical details on the recording setup see (reference  \cite{blodt2010mechanical})\\
Each signal has a length of $N = N_{a}+N_{m} = 14.10^4$, in where the first $N_a = 7.10^4$ measurement values ​​correspond to axis  in the aligned position and $N_{m} = 7.10^4$ the in misaligned position. For all the signals we used the following parameters: $d = 8$, $\tau = 1$ and $\Delta = 4.10^4$. The results are shown in Fig. \ref{fig:aJSD_wind_GIPSA}. It should be noted that for all signals, the aJSD maximum value is reached exactly when the  shaft alignment state changes. That point is identified by the dotted vertical line. In the case of the signals from the vibration measurement by the piezoelectric accelerometers as capacitive, the value of the maximum aJSD is smaller and more volatile than the other two methods. This fact can be associated with the efficacy of the measurement method. In all cases the different signals are clearly detectable by the methods here proposed.


\section{Discussion}
\label{Discussion:sec}

In paper we have introduced the notion of alphabetic Jensen Shannon Divergence to measure a
the distance between time series. The method  can be used for one-dimensional  and  two-dimensional signals. Using this measure, we have developed two methods to distinguish between different types of signals. Particularly we used them distinguish random from chaotic signals, and to compare different types of chaotic signals between them. 
Through the 
 aJSD distance matrix, we were able to show that chaotic signals are clearly distinguishable from 
 random signals with diverse power spectrum. It is important to remark that chaotic signals with different phase space structure are distinguishable from one another, also. The values taken by the entries of the distance matrix, are in correspond with the similarity (or dissimilarity of the corresponding phase spaces).
This fact was observed for  one and two dimensional chaotic sequences. Our methods were also applied to real world sequences. In particular some corresponding to ECG records, from healthy and sick patients.\\
We also developed a procedure detect dynamical changes in a signal by using a sliding window that moves on the resulting symbolic sequence after mapping the original series by a alphabetical assignment, being the centre of the original window  moved a place per step. Then the aJSD is evaluated for the two corresponding sub-sequences belonging to each half of the window.
The maximum value of the divergence  corresponds to the point where occurs a change in the probability distribution of the discretized signal. We tested the method using  two different signals
merged. The aJSD allows to detect  the  point where the two different signals were coupled in order to constitute signals with chaos-noise and chaos-chaos parts. Finally we use this scheme to evaluate the alignment condition of the axis of an electric motor.\\
Both methods have shown to be robust respect to the election of the parameters in the probability distribution evaluation. Among the main advantage we can highlight that both are easily implemented for the analysis of real world signals.



\section*{Acknowledgments}
\label{Acknowledgments:sec}

The authors thank Dr. Pierre Granjon from the GIPSA-lab, Grenoble, France,  for kindly providing them with the  recording electric motor data and  Consejo Nacional de Investigaciones Cient\'ificas y T\'ecnicas (CONICET) for financial assistance. They also thank Dr. Jose Luis Perez--Velazquez for useful discussions.

\bibliography{Chaos_Noise_bibliography}

\begin{thebibliography}{10}

\bibitem{Wold1938study}
H.~Wold.
\newblock {\em A study in the analysis of stationary time series}.
\newblock Almqvist \& Wiksell, 1938.

\bibitem{cencini2000chaos}
M.~Cencini, M.~Falcioni, E.~Olbrich, H.~Kantz, and A.~Vulpiani.
\newblock Chaos or noise: Difficulties of a distinction.
\newblock {\em Physical Review E}, 62(1):427, 2000.

\bibitem{sakai1980autocorrelations}
H.~Sakai and H.~Tokumaru.
\newblock Autocorrelations of a certain chaos.
\newblock {\em Acoustics, Speech and Signal Processing, IEEE Transactions on},
  28(5):588--590, 1980.

\bibitem{Rosso2007}
O.~A. Rosso, H.~A. Larrondo, M.~T. Martin, A.~Plastino, and M.~A. Fuentes.
\newblock Distinguishing noise from chaos.
\newblock {\em Physical review letters}, 99(15):154102, 2007.

\bibitem{zunino2012distinguishing}
L.~Zunino, M.~C. Soriano, and O.~A Rosso.
\newblock Distinguishing chaotic and stochastic dynamics from time series by
  using a multiscale symbolic approach.
\newblock {\em Physical Review E}, 86(4):046210, 2012.

\bibitem{olivares2012contrasting}
F.~Olivares, A.~Plastino, and O.~A. Rosso.
\newblock Contrasting chaos with noise via local versus global information
  quantifiers.
\newblock {\em Physics Letters A}, 376(19):1577--1583, 2012.

\bibitem{brock1986distinguishing}
W.~A. Brock.
\newblock Distinguishing random and deterministic systems: Abridged version.
\newblock {\em Journal of Economic Theory}, 40(1):168--195, 1986.

\bibitem{gao2006distinguishing}
J.~B. Gao, J.~Hu, W.~W. Tung, and Y.~H. Cao.
\newblock Distinguishing chaos from noise by scale-dependent lyapunov exponent.
\newblock {\em Physical Review E}, 74(6):066204, 2006.

\bibitem{elsner1992predicting}
J.~B. Elsner.
\newblock Predicting time series using a neural network as a method of
  distinguishing chaos from noise.
\newblock {\em Journal of Physics A: Mathematical and General}, 25(4):843,
  1992.

\bibitem{wu2004study}
Z.~Wu and N.~E. Huang.
\newblock A study of the characteristics of white noise using the empirical
  mode decomposition method.
\newblock In {\em Proceedings of the Royal Society of London A: Mathematical,
  Physical and Engineering Sciences}, volume 460, pages 1597--1611. The Royal
  Society, 2004.

\bibitem{cover2012elements}
T.~M. Cover and J.~A. Thomas.
\newblock {\em Elements of information theory}.
\newblock John Wiley \& Sons, 2012.

\bibitem{mischaikow1999construction}
K.~Mischaikow, M.~Mrozek, J.~Reiss, and A.~Szymczak.
\newblock Construction of symbolic dynamics from experimental time series.
\newblock {\em Physical Review Letters}, 82(6):1144, 1999.

\bibitem{powell1979spectral}
G.~E. Powell and I.~C. Percival.
\newblock A spectral entropy method for distinguishing regular and irregular
  motion of hamiltonian systems.
\newblock {\em Journal of Physics A}, 12(11):2053, 1979.

\bibitem{rosso2009shakespeare}
O.~A. Rosso, H.~Craig, and P.~Moscato.
\newblock Shakespeare and other english renaissance authors as characterized by
  information theory complexity quantifiers.
\newblock {\em Physica A: Statistical Mechanics and its Applications},
  388(6):916--926, 2009.

\bibitem{rosso2001wavelet}
O.~A. Rosso, S.~Blanco, J.~Yordanova, V.~Kolev, A.~Figliola, M.~Sch{\"u}rmann,
  and E.~Ba{\c{s}}ar.
\newblock Wavelet entropy: a new tool for analysis of short duration brain
  electrical signals.
\newblock {\em Journal of neuroscience methods}, 105(1):65--75, 2001.

\bibitem{Bandt2002}
C.~Bandt and B.~Pompe.
\newblock Permutation entropy: A natural complexity measure for time series.
\newblock {\em Physical Review Letters}, 88:174102, 2002.

\bibitem{yang2003linguistic}
A.~C. Yang, S.~Hseu, H.~Yien, A.~L. Goldberger, and C.~Peng.
\newblock Linguistic analysis of the human heartbeat using frequency and rank
  order statistics.
\newblock {\em Physical review letters}, 90(10):108103, 2003.

\bibitem{rao1987differential}
C.~R. Rao.
\newblock Differential metrics in probability spaces.
\newblock {\em Differential geometry in statistical inference}, 10:217--240,
  1987.

\bibitem{lin1991divergence}
J.~Lin.
\newblock Divergence measures based on the shannon entropy.
\newblock {\em Information Theory, IEEE Transactions on}, 37(1):145--151, 1991.

\bibitem{grosse2002analysis}
I.~Grosse, P.~Bernaola-Galv{\'a}n, P.~Carpena, R.~Rom{\'a}n-Rold{\'a}n,
  J.~Oliver, and H.~E. Stanley.
\newblock Analysis of symbolic sequences using the jensen-shannon divergence.
\newblock {\em Physical Review E}, 65(4):041905, 2002.

\bibitem{endres2003new}
D.~M. Endres and J.~E. Schindelin.
\newblock A new metric for probability distributions.
\newblock {\em Information Theory, IEEE Transactions on}, 49(7):1858--1860,
  2003.

\bibitem{beirlant1997nonparametric}
J.~Beirlant, E.~J Dudewicz, L.~Gy{\"o}rfi, and E.~C. Van~der Meulen.
\newblock Nonparametric entropy estimation: An overview.
\newblock {\em International Journal of Mathematical and Statistical Sciences},
  6(1):17--39, 1997.

\bibitem{robinson2011dimensions}
J.~C. Robinson.
\newblock {\em Dimensions, embeddings, and attractors}, volume 186.
\newblock Cambridge University Press, 2011.

\bibitem{takens1981detecting}
F.~Takens.
\newblock Detecting strange attractors in turbulence.
\newblock In {\em Dynamical systems and turbulence, Warwick 1980}, pages
  366--381. Springer, 1981.

\bibitem{sprott2003chaos}
J.~C. Sprott.
\newblock {\em Chaos and time-series analysis}, volume~69.
\newblock Oxford University Press Oxford, 2003.

\bibitem{knuth1973sorting}
D.~E. Knuth.
\newblock {\em Sorting and Searching, vol. 3}.
\newblock 1973.

\bibitem{steeb1992chaos}
W.~H. Steeb and M.~A. Van~Wy.
\newblock {\em Chaos and fractals: algorithms and computations}.
\newblock Wissenschaftsverlag, 1992.

\bibitem{may1976simple}
R.~M. May.
\newblock Simple mathematical models with very complicated dynamics.
\newblock {\em Nature}, 261(5560):459--467, 1976.

\bibitem{potapov2000robust}
A.~Potapov and M.~K. Ali.
\newblock Robust chaos in neural networks.
\newblock {\em Physics Letters A}, 277(6):310--322, 2000.

\bibitem{ricker1954stock}
W.~E. Ricker.
\newblock Stock and recruitment.
\newblock {\em Journal of the Fisheries Board of Canada}, 11(5):559--623, 1954.

\bibitem{arnol1961small}
V.~I. Arnol'd.
\newblock Small denominators. i. mapping the circle onto itself.
\newblock {\em Izvestiya Rossiiskoi Akademii Nauk. Seriya Matematicheskaya},
  25(1):21--86, 1961.

\bibitem{strogatz2001nonlinear}
S.~H. Strogatz.
\newblock {\em Nonlinear dynamics and chaos: with applications to physics,
  biology and chemistry}.
\newblock Perseus publishing, 2001.

\bibitem{shaw2000strange}
Shaw R. and Z.~Naturforsch.
\newblock Strange attractors, chaotic behavior, and information flow.
\newblock {\em A}, 36(80), 1981.

\bibitem{hirsch2004differential}
M.~W. Hirsch, S.~Smale, and R.~L. Devaney.
\newblock {\em Differential equations, dynamical systems, and an introduction
  to chaos}, volume~60.
\newblock Academic press, 2004.

\bibitem{henon1976two}
M.~H{\'e}non.
\newblock A two-dimensional mapping with a strange attractor.
\newblock {\em Communications in Mathematical Physics}, 50(1):69--77, 1976.

\bibitem{aronson1982bifurcations}
D.~G. Aronson, M.~A. Chory, G.~R. Hall, and R.~P. McGehee.
\newblock Bifurcations from an invariant circle for two-parameter families of
  maps of the plane: a computer-assisted study.
\newblock {\em Communications in Mathematical Physics}, 83(3):303--354, 1982.

\bibitem{nusse1998dynamics}
H.~E. Nusse.
\newblock {\em Dynamics: numerical explorations}, volume 101.
\newblock Springer, 1998.

\bibitem{schmidt1985dissipative}
G.~Schmidt and B.~W. Wang.
\newblock Dissipative standard map.
\newblock {\em Physical Review A}, 32(5):2994, 1985.

\bibitem{arnold1968ergodic}
V.~I. Arnolʹd and A.~Avez.
\newblock {\em Ergodic problems of classical mechanics}, volume~9.
\newblock Benjamin, 1968.

\bibitem{chernikov1988chaos}
A.~A. Chernikov, R.~Z. Sagdeev, and G.~M. Zaslavskii.
\newblock Chaos-how regular can it be?
\newblock {\em Physics Today}, 41:27--35, 1988.

\bibitem{chirikov1979universal}
B.~V. Chirikov.
\newblock A universal instability of many-dimensional oscillator systems.
\newblock {\em Physics reports}, 52(5):263--379, 1979.

\bibitem{devaney1984piecewise}
R.~L. Devaney.
\newblock A piecewise linear model for the zones of instability of an
  area-preserving map.
\newblock {\em Physica D: Nonlinear Phenomena}, 10(3):387--393, 1984.

\bibitem{LarrondoProg}
Larrondo H.A.
\newblock program: noisef k.m,
  http://www.mathworks.com/matlabcentral/fileexchange/35381, 2012.

\bibitem{rabin1980probabilistic}
M.~O. Rabin.
\newblock Probabilistic algorithm for testing primality.
\newblock {\em Journal of number theory}, 12(1):128--138, 1980.

\bibitem{blodt2010mechanical}
M.~Bl{\"o}dt, P.~Granjon, B.~Raison, and J.~Regnier.
\newblock Mechanical fault detection in induction motor drives through stator
  current monitoring-theory and application examples.
\newblock {\em Fault Detection}, pages 451--488, 2010.

\end{thebibliography}
\bibliographystyle{unsrt}
\end{document}